\documentclass[twocolumn,preprintnumbers,amsmath,amssymb,floatfix,showpacs,prl,aps]{revtex4}

\usepackage{dcolumn}
\usepackage{bm}
\usepackage{graphicx}

\begin{document}


\title{
A Geometrical Description of the ``Classical'' Weak-Field 
Hall Effect in Metals
}
\author{F. D. M. Haldane}
\affiliation{Department of Physics, Princeton University,
Princeton NJ 08544-0708}

\date{April 12, 2005}

\begin{abstract}
Within the ``anisotropic relaxation time'' approach to transport processes,
I obtain a new  geometric  formula for the
weak-field Hall conductivity of metals, 
which generalizes previously-known formulas 
due to Tsuji (restricted to cubic 3D metals) and Ong (for 2D metals).
\end{abstract}


\maketitle

While the Kubo formula in principle
provides a fully-Hamiltonian description of
the conductivity of metals, practical
calculations generally  
use the simpler  linearized semiclassical 
Boltzmann equation description, commonly
combined with the additional ``relaxation time'' approximation\cite{jz}.
This leads to  a  ``geometric'' expression for the
DC Ohmic conductivities, which can be
expressed in terms of the Fermi-surface geometry and
a temperature-dependent
 ``scattering path length'' (mean free path) $\ell(\bm s,T)$ =
$v_F(\bm s)\tau(\bm s,T)$ where $\bm s$ = $\{s^1,s^2\}$ is a
pair of curvilinear coordinates 
parameterizing points on the Fermi surface $\bm k_F(\bm s)$,
$v_F(\bm s)$ is the Fermi speed, and $\tau(\bm s,T)$ is a notional
temperature-dependent ``anisotropic'' ($\bm s$-dependent) relaxation
time for the quasiparticle distribution to relax back into equilibrium
once the current-driving field is removed. 

The phenomenological ``scattering
path length'' can also be expressed as a vector mean free path,
$\bm \ell(\bm s)$ = $\ell(\bm s)\hat{\bm n}_F(\bm s)$, where
$\hat {\bm n}_F(\bm s)$ is the outward normal of the Fermi surface
(direction of the Fermi velocity).   
The detailed microscopic origin of $\ell(\bm s,T)$ is usually rather 
inaccessible: theoretically, within the Boltzmann formalism, 
one must distinguish
between ``quasiparticle lifetimes'', and 
``transport lifetimes'',
but the uncertainties of the microscopic relaxation mechanisms turn
the scattering path length into a phenomenological fitting parameter.
(See Ref.\cite{allenreview} for a review of the application
of Boltzmann theory to metals.)
It is probably most meaningful in the case where the dominant
relaxation process is ``catastrophic'' inelastic scattering, so that
all memory of the previous quasiparticle  state is erased by the scattering process.

In this formalism, with unbroken time-reversal symmetry,
the  symmetric DC Ohmic conductivity tensor is
\begin{equation}
\sigma^{ab}_{\Omega}(T) 
= \frac{e^2}{\hbar}
\sum_{\alpha} g_s \int_{\mathcal S_{\alpha}}\frac{d^2S_F}
{(2\pi)^3}
\hat{n}_F^a(\bm s)\hat{n}_F^b(\bm s) \ell(\bm s,T) ,
\label{jz0}
\end{equation}
where $g_s$ = 2 is the spin degeneracy factor, $\mathcal S_{\alpha}$ are
the topologically-distinct sheets (compact 2-manifolds) of the Fermi surface
(labeled by $\alpha$), and $d^2S_F$ is the (scalar) Fermi-surface area
element.
To make the structure of geometrical quantities clearer, I 
use a ``covariant'' notation where spatial coordinates $r^a$,
$a$ = 1,2,3, have
``upper'' indices, and the dual $k$-space coordinates $k_a$ have ``lower''
indices, with the summation convention on repeated upper-lower pairs; this
shows up the formal difference  between $\bm k\cdot \bm r$
= $k_a r^a$ which is metric-independent, and  
$\hat{\bm n}_F\cdot \hat{\bm n}_F$
= $g^0_{ab}\hat n_F^a \hat n_F^b$ = 1, 
where $g^0_{ab}$ = ``$\delta_{ab}$'' =
${\rm diag}(1,1,1)$ is the Euclidean metric tensor with
$\det g^0$ = 1.  
In such a notation, the
components of the electric field $E_a$ have lower indices, while
components $B^a$ of the magnetic flux density have upper indices.
In a metal with cubic symmetry,  the conductivity tensor is isotropic,
$\sigma_{\Omega}^{ab}(T)$ $\propto$ $g_0^{ab}$, where
$g_0^{ab}$ is the
inverse of the Euclidean metric
tensor:
\begin{equation}
\sigma^{ab}_{\Omega}
= 
{\textstyle \frac{1}{3}}
g_0^{ab}
\frac{e^2}{\hbar}
\sum_{\alpha} g_s
\int_{\mathcal S_{\alpha}}\frac{d^2S_F}{(2\pi)^3}
\ell(\bm s,T) ,
\quad \mbox{(cubic)}.
\label{cube1}
\end{equation}

If time-reversal symmetry is broken by the application of a weak uniform
magnetic flux density $\bm B$, $\sigma^{ab}$
acquires an antisymmetric Hall term $\sigma^{ab}_H$:  
to linear order in $\bm B$, 
\begin{equation}
\sigma^{ab}(T,\bm B)=
\sigma^{ab}_{\Omega}(T) + \frac{e^2}{\hbar}\epsilon^{abc}\gamma_{cd}(T)
\left (\frac{eB^d}{\hbar}\right ),
\end{equation}
where $\epsilon^{abc}$ is the usual  antisymmetric rank-3 Levi-Civita
symbol, and $\gamma_{ab}(T)$ is a symmetric tensor.
Using the same Jones-Zener solution\cite{jz} of
the Boltzmann equation in the relaxation-time
approximation that is usually used to derive
(\ref{jz0}), Tsuji\cite{tsuji} found a geometric expression for the
weak-field Hall conductivity of cubic metals, which is equivalent to
\begin{equation}
\gamma_{ab} = 
{\textstyle \frac{1}{3}}
g^0_{ab}
\sum_{\alpha} g_s  \int_{\mathcal S_{\alpha}}\frac{d^2S_F}{(2\pi)^3}
H(\bm s) \ell^2(\bm s,T) ,
\quad \mbox{(cubic)},
\label{cube2}
\end{equation}
where $H(\bm s)$ is the \textit{mean curvature} of the Fermi surface
at point $\bm s$.    
The Tsuji formula has been quantitatively tested\cite{schultz92} 
against band-structure calculations of the Fermi surface
curvatures of various cubic metals, plus ``reasonable'' assumptions
about $\tau(\bm s,T)$ (and hence $\ell(\bm s,T)$); that study indeed found
sensitivity to Fermi surface regions of high mean curvature.

Unlike (\ref{jz0}), the formulas (\ref{cube1}) and (\ref{cube2}) are not truly ``local''
Fermi surface formulas, 
because they essentially perform an 
an integral of the average over 
the ``star'' of 48 equivalent points of a general Fermi surface point
in a system with cubic symmetry.   The purpose of this Letter is to give the
full non-cubic generalization of Tsuji's formula as a true integral over local Fermi-surface properties; this new
 ``classical''
Hall effect formula  complements another new geometric 
Fermi-surface formula\cite{fdmh} for the intrinsic 
``anomalous'' ($\bm B$ = 0) Hall conductivity of ferromagnetic metals,
which is a 
fundamental quantum effect associated with quasiparticle Berry phases.

The Fermi-surface curvature  field $\kappa^{ab}(\bm s)$ is a symmetric tensor
relating two ``one-forms'', the differential change $d\hat{\bm n}_F$
in outward normal vector and the differential change $d\bm k_F$ of the
Fermi vector for small
displacements  $d \bm s$ on the Fermi surface:
\begin{equation}
d\hat{n}_F^a = \kappa^{ab}(\bm s) dk_{Fb} .
\label{curv}
\end{equation}
In a Euclidean coordinate frame that locally diagonalizes it at Fermi surface
point $\bm s$,
$\kappa^{ab}(\bm s)$  =${\rm diag}(k_1^{-1},k_2^{-1},0)$,
and $n_F^a(\bm s)n_F^b(\bm s)$ = 
${\rm diag}(0,0,1)$, where $k_1$ and $k_2$ are the two principal
Fermi-surface radii of curvature.  Then
\begin{equation}
H(\bm s) = \frac{1}{2}\left (\frac{1}{k_1}+\frac{1}{k_2}\right ),
\quad K(\bm s) = \frac{1}{k_1k_2},
\end{equation}
are respectively the mean (extrinsic) curvature and the
Gaussian (intrinsic) curvature.   Note that
$(H(\bm s))^2$ $\ge$ $K(\bm s)$, so the mean curvature can only change sign
(from electron-like  $H(\bm s) > 0$ to
hole-like $H(\bm s) < 0$) in regions of negative Gaussian
curvature.   The relation (\ref{curv}) can be also be expressed as
\begin{equation}
dk_{Fa} 
= k_{ab}(\bm s) 
d\hat{n}_F^b, 
\end{equation}
where $k_{ab}(\bm s)$ = ${\rm diag}(k_1,k_2,0)$ is the
\textit{radius-of-curvature field}, which is a well-defined
symmetric tensor  wherever
the Gaussian curvature $K(\bm s)$ is non-zero.
It diverges when $K(\bm s)$ $\rightarrow$ 0,
but the product $K(\bm s)k_{ab}(\bm s)$ 
$\equiv$ $\epsilon_{ace}\epsilon_{bdf}\hat n_F^c(\bm s)
\hat n_F^d(\bm s)\kappa^{ef}(\bm s)$
remains finite.

The Gaussian curvature $K(\bm s)$ has a profound geometric significance;
it relates two fundamental ``two-forms'', the Fermi surface (scalar)
area element $d^2S_F$ = $g^0_{ab}\hat n_F^a\epsilon^{bcd}
dk_{Fc}\wedge dk_{Fd}$,
and $d^2\Omega_F$ = $\epsilon_{abc}\hat n_F^a d\hat n_F^b\wedge d\hat n_F^c$, 
the solid angle subtended by the
normal vector $\hat{\bm n}_F(\bm s)$ as it varies over that area element:
\begin{equation}
K(\bm s)d^2S_F = d^2\Omega_F .
\end{equation}
This leads to the Gauss-Bonnet theorem that
\begin{equation}
\frac{1}{4\pi}\int_{S_{\alpha}} K(\bm s) d^2S_F 
= 
\frac{1}{4\pi}\int_{S_{\alpha}}d^2\Omega_F 
= 1-g_{\alpha},
\end{equation}
where the non-negative integer $g_{\alpha}$ is the \textit{genus}
(number of ``handles'') of a Fermi-surface manifold $\mathcal S_\alpha$.

The genus is the primary topological characteristic
of a Fermi surface manifold; another important topological property is the 
the dimension $d^G_{\alpha}$ of the Bravais Lattice of reciprocal vectors
$\bm G$ generated by periodic open orbits on $\mathcal S_{\alpha}$.
For a closed path $\Gamma$,
$\oint_{\Gamma} d\bm k_F$ = $\bm G(\Gamma)$,
where $\bm G(\Gamma)$ is a non-zero reciprocal lattice vector if the image
of $\Gamma$ in 3D $k$-space is a periodic ``open orbit''.   
On a 2-manifold of
genus $g_{\alpha}$, all closed paths $\Gamma$ are topologically
equivalent to paths built up as sequences of paths chosen
from a set of  $2g_{\alpha}$ elementary
non-trivial closed paths $\Gamma_i$ (the basis of
 generators of the homology group).
The generators $\Gamma_i$ can be chosen
so that only $d^G_{\alpha} \le 3 $ of them
have non-zero $\bm G(\Gamma_i)$, which are
linearly independent, and form a basis of
a $d^G_{\alpha}$-dimensional Bravais lattice. A 
sphere-like ``nearly-free-electron'' 3D Fermi surface,
as in the alkali metals, has $d^G_{\alpha}$ = 0, a
cylinder-like  quasi-2D Fermi
surface has $d^G_{\alpha}$ = 1, a plane-like quasi-1D Fermi surface
has $d^G_{\alpha}$ = 2, and a multiply-connected Fermi surface with
cubic symmetry, as in the noble metals, has $d^G_{\alpha}$ = 3.

My main result,
the non-cubic generalization of the Tsuji formula, can now be given:
\begin{equation}
\gamma_{ab}(T) = 
\sum_{\alpha}
\frac{g_s}{2}
  \int_{\mathcal S_{\alpha}}
\frac{K(\bm s)d^2S_F}{(2\pi)^3}
k_{ab}(\bm s) \ell^2(\bm s,T) .
\label{tsuji2}
\end{equation}
Unlike (\ref{cube2}), this is a true integral of local quantities
on the Fermi surface, making no use of any point-group symmetries,
and is expressed in a fully-covariant tensor formulation, independent of
choice of  coordinate frame.   In the cubic case, the replacement
$\gamma_{ab}$ $\rightarrow$ $\frac{1}{3}g^0_{ab}g_0^{cd}\gamma_{cd}$ may be 
made, and the cubic Tsuji form
(\ref{cube2}) is recovered because $g_0^{ab}k_{ab}(\bm s)$ 
$\equiv$ $2H(\bm s)/K(\bm s)$.

The derivation of (\ref{tsuji2}) is as follows; let 
$n_0(\bm k)$, for $\bm k$ near the Fermi surface in 3D $k$-space, be
the ground-state Fermi-Dirac distribution of quasiparticle occupations
in an independent-quasiparticle Ansatz, and let $\delta n_E(\bm k)$
= $(e\tau(\bm k)/\hbar)\bm E \cdot \bm \nabla_k n_0(\bm k)$ be
the first-order non-equilibrium change induced by the uniform
applied electric field $\bm E$, where $\bm \nabla_k$ is the $k$-space
derivative, and $\tau(\bm k)$ is an appropriate phenomenological
temperature-dependent anisotropic relaxation time.
Here $\bm \nabla_k n_0(\bm k)$ = $\hbar \bm v_F(\bm k) 
f'(\delta \varepsilon(\bm k))$, where $f(\varepsilon)$
is the Fermi-Dirac distribution function 
$(\exp (\varepsilon /k_BT) + 1)^{-1}$, 
$\delta \varepsilon (\bm k)$
is the quasiparticle excitation energy,
and $\bm v_F(\bm k)$ = $\hbar^{-1}\bm \nabla_k \delta 
\varepsilon (\bm k)$ is the quasiparticle group velocity.
Then
\begin{equation}
\delta n_E(\bm k) = 
f'(\delta \varepsilon(\bm k))e\bm E \cdot \bm \ell,
\end{equation}
where $\bm v_F(\bm k)\tau(\bm k)$ $\equiv$ $\bm \ell (\bm k)$ $\equiv$ 
$\ell(\bm k)\hat{\bm n}_F$.
Because it induces $k$-space motion parallel to the Fermi surface,
a weak uniform magnetic flux density only produces a second-order change
$\delta n_{E,B}(\bm k)$ = 
$(e\bm v_F(\bm k)\tau(\bm k)/\hbar)\times \bm B \cdot \bm \nabla_k
\delta n_E(\bm k)$:
\begin{equation}
\delta n_{E,B}(\bm k) =
f'(\delta \varepsilon(\bm k))\bm \ell \times (e\bm B/\hbar) 
\cdot \bm \nabla_k (e\bm E\cdot \bm \ell).
\end{equation}

The current density is
$g_s V^{-1}\sum_k e\bm v_F(\bm k) \delta n(\bm k)$, where $V$ is
the volume of the system;
in the Fermi-liquid regime, $f'(\delta \varepsilon (\bm k))$ vanishes
except near the Fermi surface, and the expression for
the induced current becomes a Fermi surface integral.
The first-order term $\delta n_E$ gives the Ohmic conductivity
$\sigma^{ab}_{\Omega}$
(\ref{jz0}); $\delta n_{E,B}$ gives the Hall term $\sigma^{ab}_H$ as
\begin{equation*}
\frac{e^3}{\hbar^2}\sum_{\alpha}
g_s \int_{\mathcal S_{\alpha}} \frac{d^2S_F}{(2\pi)^3}
( \hat{n}_F^a\ell) \left (\bm B \times 
\hat{\bm n}_F \cdot \bm \nabla_k\right ) ( \hat n_F^b \ell). 
\end{equation*}
This is an  antisymmetric tensor in indices $a$, $b$,
as 
\begin{equation*}
d^2S_F (\bm B \times \hat{\bm n}_F\cdot \bm \nabla_k)
\equiv Bdk_{F\parallel}dk_{F\perp}\partial/\partial k_{F\perp},
\end{equation*}
where $dk_{F\parallel}$ and $dk_{F\perp}$ are components of
$d\bm k_F$ parallel and perpendicular to $\bm B$; the symmetrized
form of the expression for $\sigma^{ab}_H$
is  the integral 
of the derivative of a periodic function
over a full period, and thus vanishes. 
After explicitly antisymmetrizing the expression, 
one can replace
$\bm \nabla_k (\hat n_F^b\ell)$ by
$\ell (\bm \nabla_k \hat n_F^b)$, and
then $\bm B \times \hat{\bm n}_F \cdot \bm \nabla_k \hat n_F^b$ 
by
$\epsilon_{cde}B^c\hat n_F^d\kappa^{eb}$; finally the identity
$\epsilon_{ace}\epsilon_{bdf}\hat n_F^c\hat n_F^d\kappa^{ef}$
$\equiv$ $Kk_{ab}$ is used to obtain (\ref{tsuji2}).

There have been previous attempts to generalize the Tsuji formula
(\textit{e.g.}, Ref.\cite{kesternich}); these did not uncover a
general geometric formula with the simplicity of (\ref{tsuji2}) because
they typically expressed the solution of the Boltzmann equation in terms
of non-geometric quantities such as higher $k$-space derivatives of
the quasiparticle dispersion $\delta \varepsilon(\bm k)$. What is remarkable
about (\ref{tsuji2}) (and (\ref{jz0})) is that they only involve Fermi surface
geometry and the scattering path length, with no direct 
reference to Fermi-liquid energy parameters such as $v_F(\bm s)$. 
As in general relativity, the use of a covariant metric-independent
and coordinate-frame-independent tensor formulation greatly aids
the identification of formulas with the correct geometric structure.  

There is a subtle change of the interpretation of the Fermi surface
in the formulas presented here.   Instead of thinking of the Fermi surface
as a set of  2D surfaces embedded in a  periodically-repeated
3D  $k$-space, it is best to view it
as a set of compact 2-manifolds $\mathcal S_{\alpha}$
on which quasiparticle states are defined, and
which are the domains of functions such as $\bm k_F(\bm s)$,
$\hat{\bm n}_F(\bm s)$, $k_{ab}(\bm s)$, \textit{etc}.    
On such a 2-manifold, \textit{any} ``two-form'' can be used to 
formulate a surface integral.
In particular, $\ell^2 K d^2S_F $ = $\ell^2 d^2\Omega_F$ =
$\ell^{-1}\bm \ell \cdot d^2\bm S_{\ell}$, where
$(d^2  S_{\ell })_a$ is the (oriented) \textit{vector} surface area 
element
$\epsilon_{abc} d\ell^b\wedge d\ell^c $
of the (temperature-dependent) 2D
surface swept out by $\bm \ell(\bm s,T)$ as
$\bm s$ varies over $\mathcal S_{\alpha}$.

This leads to an alternative form of (\ref{tsuji2}),
inspired by the novel approach  of  Ong\cite{ong}, who
studied the Hall conductivity $\sigma^{2D}_H$  of a 2D metal
where quasiparticle motion 
takes place only on electronically-isolated 
lattice planes 
with no interplane tunneling,  and where
the Fermi surface is a set of 1-manifolds $\Gamma_{\alpha}$, so
$\bm k_F(s)$
is a  one-parameter curve in the planar 
Brillouin zone.
Ong pointed out that, instead of  focusing 
on $\bm k_F(s)$, it was far more informative to
examine the planar curve  traced out by
the 2D mean-free-path vector
$\bm \ell_0(s,T)$, and obtained 
\begin{equation}
\sigma^{\rm 2D}_H(T) = 
\frac{e^2}{\hbar}
\sum_{\alpha} 
 \frac{g_s}{2}
\oint_{\Gamma_{\alpha}}
\frac{d \bm \ell_0}{(2\pi)^2} \cdot 
\bm \ell_0 \times 
\frac{e\bm B}{\hbar},
\label{ongform}
\end{equation}
which
expresses
the 2D Hall conductivity of a single  plane 
\textit{completely} in terms of $\bm \ell_0(s,T)$.
In analogy to Ong's approach, the 3D formula (\ref{tsuji2}) 
can also  be
written as
\begin{equation}
\gamma_{ab}(T) = 
\sum_{\alpha}
\frac{g_s}{2} 
 \int_{\mathcal S_{\alpha}}\frac{\bm \ell \cdot d^2 
\bm S_{\ell}}
{(2\pi)^3}\,
\,
 k_{ab}(\bm s) \ell^{-1}(\bm s) .
\label{tsuji2a}
\end{equation}

\begin{figure}
\includegraphics[width=3in]{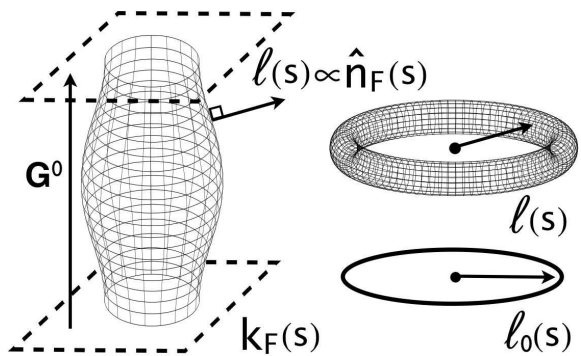}
\caption{\label{fig1}
(Left) 3D  $k$-space image (Fermi vector $\bm k_F(\bm s)$),
 and (right) the corresponding 3D 
$\ell$-space image (vector mean-free-path $\bm \ell (\bm s)$)
of a quasi-2D Fermi surface (quasiparticle 2-manifold parametrized
by $\bm s$).   $\bm G^0$
is the primitive reciprocal vector that indexes the weakly-coupled
lattice planes.  
$\bm \ell (\bm s)$ is parallel to the outward
Fermi-surface normal $\hat{\bm n}_F(\bm s)$.
Both images are of a surface with genus $g$ = 1, an
intrinsic property independent of the coordinates ($\bm k_F(\bm s)$
or $\bm \ell(\bm s)$) used to represent the 2-manifold.
In the 2D limit, the toroid $\bm \ell (\bm s)$ collapses into a 
planar loop $\bm \ell_0( s)$ with $\bm G^0\cdot \bm \ell_0$ = 0.
}
\end{figure}

If a 3D system is formed by a family of isolated 
metallic 2D lattice planes
indexed by a primitive reciprocal lattice vector $\bm G^0$, 
$\sigma^{ab}_H$ = $\sigma^{\rm 2D}_H\epsilon^{abc}(G^0_c/2\pi)$,
with  $\bm G^0 \cdot \bm \ell_0 $ = 0. Then the 2D
formula (\ref{ongform}) is
equivalent to
\begin{equation}
\gamma_{ab}(T) = 
G^0_a\epsilon_{bcd}
\sum_{\alpha} 
\frac{g_s}{2}
\oint_{\Gamma_{\alpha}}
\frac{d\ell_0^c}{(2\pi)^3}
\ell_0^d ,
\quad \mbox{(2D limit)}.
\label{ongeq}
\end{equation}
Then (\ref{ongeq})  can be obtained from (\ref{tsuji2a}) 
in the limit that a quasi-2D Fermi surface
with genus $g$ =  1 (and $d^G$ = 1)
degenerates into
a strictly-2D system (see Fig.\ref{fig1}). 
The limiting process is not entirely straightforward, 
because in the 
2D limit, the Gaussian curvature $K(\bm s)$ $\rightarrow $ 0 everywhere
on the Fermi surface.  The $k$-space image of such a Fermi surface is
a weakly-modulated cylindrical surface connecting two opposite
Brillouin-zone faces (relatively-displaced by
$\bm G^0$), while the $\ell$-space image is a 
toroid, with one small radius that shrinks to zero in the 2D limit,
so the toroid collapses into 
a loop  $\bm \ell_0(s)$ in the plane  $\bm \ell_0 \cdot \bm G^0$ = 0.

For passage to the 2D limit,
the vector mean-free-path $\bm \ell (\bm s)$ can be written as 
$\bm \ell_0(s) + \delta \bm \ell (s,s_{\perp})$, where 
$\bm \ell_0\cdot \bm G^0$ = 0, and
$\delta \bm \ell \cdot \bm G^0$ depends only on $s_{\perp}$;
here, both $s$ and $s_{\perp}$ are periodic parametric
Fermi surface coordinates, so displacements $d\bm k_F$
with  constant $s_{\perp}$ 
are normal
to $\bm G^0$, and those with constant $s$ are along the component
of $\bm G^0$ parallel to the Fermi surface.
In the 2D limit,
$\delta\bm \ell (\bm s)$ $\rightarrow 0$,
but $k_{ab}(\bm s)$ has a divergent part $\Lambda(\bm s) G^0_aG^0_b$,
where $\Lambda \rightarrow \infty$.
Any path along which $s_{\perp}$ increases at fixed $s$ is a
periodic open orbit with
$\oint ds_{\perp} k_{ab}\partial n^b_F/\partial s_{\perp}$ = $\bm G^0_a$;
so as $\delta \bm \ell \rightarrow 0$,
$\oint ds_{\perp} \Lambda (\partial/\partial s_{\perp})
 \delta \bm \ell \cdot \bm G^0 $
$\rightarrow$ $\ell_0(s)$, which remains finite.
In the 2D limit,
$\bm \ell \cdot d^2\bm S_{\ell}  $ $\rightarrow$ 
$\left (\bm \ell_0 \times d\bm \ell_0 \right )
\cdot 
(\partial  \delta \bm \ell /\partial s_{\perp})  ds_{\perp}$.
Since $\bm \ell_0 \times d\bm \ell_0$ is parallel to $\bm G^0$,
(\ref{tsuji2a}) is now easily found to evolve into (\ref{ongeq}).

In 2D, $\sigma^{\rm 2D}_H$ has dimensions $[\mbox{resistance}]^{-1}$, and
is  $e^2/\hbar$ times a dimensionless factor that is invariant
under rescaling of the Fermi surface.  In 3D, in contrast, 
$\sigma^{ab}_H$ has dimensions 
$[\mbox{resistance}]^{-1}\cdot [\mbox{wavenumber}]$
and scales linearly with Fermi-surface scale; thus the 3D
formula
(\ref{tsuji2a}) depends on Fermi-surface geometry
through the radius-of-curvature field
$k_{ab}(\bm s)$, and not just  on the vector mean-free-path
$\bm \ell (\bm s,T)$ alone, unlike
the 2D formula (\ref{ongform}).

There is an important difference between the images of the
Fermi surface 2-manifold as represented by the Fermi vector
field  $\bm k_F(\bm s)$ and
the vector mean-free-path field $\bm \ell (\bm s)$. 
In $k$-space, ``level repulsion'' means that
the Fermi surface $\bm k_F(\bm s)$
is generically non-self-intersecting (except
in exceptional non-symmorphic space-group
cases where it crosses  certain Brillouin zone
boundaries).   Even point degeneracies, associated with
intersections with axes of $k$-space rotation symmetry (or
rare stable ``accidental'' degeneracies), are not common.    In constrast,
$\ell$-space surfaces $\bm \ell (\bm s)$ have no self-avoidance property,
and may well intersect in complicated ways.

The formulas (\ref{tsuji2}) and (\ref{tsuji2a}) presented here are
``weak field'' (small-$\bm B$) formulas.   The ``weak-field'' condition
is usually characterized  as ``$\omega_c\tau \ll 1$''.  The  appropriate
geometrical condition is that the Fermi-surface displacement
$\Delta \bm k_F$ = $\bm \ell(\bm s) \times (e\bm B/\hbar)$
as  the quasiparticle travels its mean free path
is such that the change
$\Delta \hat{\bm n}_F$  in its real-space 
direction-of-motion is small, where
\begin{equation}
\left (\hat{\bm n}_F \times \Delta \hat{\bm n}_F \right )_a
= 
\ell(\bm s,T)K(\bm s) k_{ab}(\bm s) \left ( \frac{eB^b}{\hbar} \right ).
\end{equation}

From a geometric 
Fermi-liquid point of view, the scattering path length $\ell (\bm s,T)$
may be a ``more fundamental'' quantity than the quasiparticle
lifetime $\tau(\bm s,T)$.  It controls the maximum possible real-space extent 
of a Gaussian quasiparticle wavepacket, as coherence would be lost if the
size was larger.   By the uncertainty principle, $\ell(\bm s,T)^{-1}$ 
is a lower bound on the quasiparticle localization near the Fermi surface
in $k$-space.   Assuming that ``natural''
Gaussian quasiparticle
wavepackets should have quasi-isotropic spreads in $k$-space,
this local $k$-space lengthscale may also provide a natural
``quantum geometric'' scale for non-orthogonality of wavepackets centered at
neighboring Fermi surface points, and may play the role of
a (temperature-dependent) ``quantum length'' analogous  to the magnetic 
length $\ell_B$ = $\surd (\hbar/|eB|)$ in lowest-Landau level physics.

It  seems appropriate to characterize a transport regime in which the
new weak-field Hall conductivity formulas (\ref{tsuji2}), (\ref{tsuji2a})
are  valid as a ``classical'' Hall regime, as it seems to involve no
aspect of the quasiparticle dynamics other than the underlying geometrical
constraint provided by the Fermi surface, which defines a
(3+2)-dimensional symplectic phase-space structure that, at each $\bm s$, 
singles out one
real-space direction ($\hat{\bm n}_F(\bm s)$) as ``special'' (not conjugate
to a displacement $d\bm k_F$  on the Fermi surface).  Equilibrium fluctuations
driven by ``catastrophic'' inelastic scattering events nucleate a quasiparticle
at Fermi surface point $\bm s$, and it propagates a distance $\ell(\bm s,T)$
in the local ``special'' direction before another violent inelastic event
causes its decoherence, and returns the charge to the local ``reservoir''.
During its lifetime, the quasiparticle's trajectory can be influenced by
electric and magnetic fields.
This seems to be an entirely ``classical'' diffusive process, with no
involvement of quantum interference or Hamiltonian dynamics, as none
of the microscopic ingredients of Fermi liquid theory (Fermi speed 
$v_F(\bm s)$, Landau  functions, 
quasiparticle Berry phases\cite{fdmh}, \textit{etc}.) are referenced in
(\ref{tsuji2}), (\ref{tsuji2a}).

In summary, I have described a geometric formulation  of the
``classical'' weak-field Hall effect in metals (within
a relaxation-time approach),
and given two equivalent
forms 
(\ref{tsuji2}), (\ref{tsuji2a})
of a new  formula expressing the Hall conductivity
entirely in terms of Fermi surface geometry
and a quasiparticle mean-free-path length
that is traversed between scattering events
that erase memory of its prior history.    This  generalizes previously-known
formulas that were restricted to special cases: Tsuji's formula\cite{tsuji}
for metals with cubic symmetry, and Ong's formula\cite{ong} for 2D metals.

This work was supported in part by the U. S. National
Science Foundation (under MRSEC Grant No. DMR02-13706) at  the Princeton
Center for Complex Materials.

\end{document}